\documentclass[a4paper,11pt]{article}
\usepackage{jheppub} 

\usepackage{xcolor}
\usepackage{graphicx}
\usepackage{dcolumn}
\usepackage{bm}
\usepackage{hyperref}
\usepackage{multirow}
\usepackage{slashed}

\title{QCD Factorization from Light-ray OPE}

\author[]{Hao Chen}

\affiliation[]{Zhejiang Institute of Modern Physics, School of Physics, Zhejiang University, Hangzhou, 310027, China}

 \emailAdd{chenhao201224@zju.edu.cn}

\abstract{The energy-energy correlator~(EEC) in Quantum Chromodynamics~(QCD) serves as an important event shape for probing the substructure of jets in high-energy collisions. A significant progress has been make in understanding the collinear limit, where the angle between two detectors approaches zero, from the factorization formula in QCD and the light-ray Operator Product Expansion~(OPE) in Conformal Field Theory. Building upon prior research on the renormalization of light-ray operators, we take an innovative step to extend the light-ray OPE into non-conformal contexts, with a specific emphasis on perturbative QCD. Our proposed form of the light-ray OPE is constrained by three fundamental properties: Lorentz symmetry, renormalization group invariance, and constraints from physical observables. This extension allows us to derive a factorization formula for the collinear limit of EEC, facilitating the future exploration and understanding on subleading power corrections in collinear limit.}

\begin{document}
\maketitle
\flushbottom


\section{Introduction}

Jets are the consequence of real-time dynamics of Quantum Chromodynamics~(QCD) in the high-energy collisions. They are collimated sprays of hadrons, originated from fragmentation of energetic partons produced in hard scatterings~\cite{Sterman:1977wj,Salam:2010nqg}. Event shape observables provide a window to study the QCD jet dynamics, furthering our understanding of Lorentzian quantum field theory. The energy-energy correlation~(EEC) is one of the most interesting infrared-safe observables~\cite{Basham:1978bw,Basham:1978zq}. In recent years, EEC has been attracting interest from theoretical, phenomenological and experimental sides, especially in the context of collinear limit~\cite{Hofman:2008ar,Dixon:2019uzg,Kologlu:2019mfz,Chen:2020vvp,Komiske:2022enw,CMS:2023wcp,Tamis:2023guc}. 

The function of EEC is to measure the correlation of the energy flow into two detectors separated by an angle $\theta$ on the celestial sphere
\begin{equation}
    \mathrm{EEC}(z)=\frac{1}{\sigma}\sum_{i,j}\int d\sigma \frac{E_i E_j}{Q^2} \delta(z-\frac{1-\cos\theta_{ij}}{2})\,,\label{eq:EEC_def_1}
\end{equation}
where the distribution is normalized to the total cross section $\sigma$. $d\sigma$ is the differential cross section, $E_i$ is the energy of the particle $i$, $Q$ is the center of mass energy in the $e^+ e^-$ annihilation or the Higgs decay, $\theta_{ij}$ is the angle between the particles $i,j$, and $z=(1-\cos\theta)/2$ is related to the angle $\theta$ between the two detectors. On the other hand, EEC can be defined as the correlation function of the energy flow operators~\cite{Hofman:2008ar}
\begin{equation}
    \mathrm{EEC}(z)= \frac{8\pi^2}{Q^2}\langle\mathcal{E}(n)\mathcal{E}(n^\prime)\rangle
     =\frac{8\pi^2}{q^2 \sigma} \! \int \! d^4 x\, e^{i q \cdot x} \langle\Omega|    \mathcal{O}^\dagger(x)  {\cal E}(n) {\cal E}(n^\prime) \mathcal{O}(0)|\Omega \rangle \,,\label{eq:EEC_def_2}
\end{equation}
where null vectors $n = (1, \vec{n})$ and $n^\prime = (1, \vec{n}^\prime)$ describe the directions of the detectors and relate to $z$ through $z=n\cdot n^\prime/2$, $q^\mu=(Q,\vec{0})$ is the total momentum, $\mathcal{O}$ is the source creates the excitations on the vacuum $|\Omega\rangle$, and the energy flow operator $\mathcal{E}(n)$ is an integral transformation of the local stress tensor $T^{\mu\nu}$~\cite{Sveshnikov:1995vi}
\begin{equation}
    \mathcal{E}(n) = \lim_{r\to\infty} r^2 \int_{-\infty}^{\infty} dt\, \vec{n}^i T^0_{\; i}(t,r\vec{n}) \,. \label{eq:energy_flow_operator}
\end{equation}
The two equivalent definitions \eqref{eq:EEC_def_1} and \eqref{eq:EEC_def_2}, from momentum space and position space perspectives respectively, make EEC an interesting observable which benefits both from scattering amplitude and correlation function techniques. Motivated by these definitions, there are a number of interesting generalizations of the EEC, including the multi-point and projective versions~\cite{Chen:2020vvp,Chen:2019bpb,Yang:2022tgm,Yan:2022cye,Chen:2020adz,Chen:2021gdk}, track-based EEC~\cite{Chen:2020vvp,Li:2021zcf,Jaarsma:2023ell,Lee:2023tkr,Lee:2023xzv}, Transverse EEC at hadron colliders~\cite{Ali:1984yp}, generalized event shapes~\cite{Korchemsky:2021okt,Korchemsky:2021htm}, nuclear EEC~\cite{Liu:2022wop,Liu:2023aqb,Cao:2023oef,Li:2023gkh}, the celestial non-gaussianities~\cite{Chen:2022swd}, and in various backgrounds like quark-gluon plasma and cold nuclear matter~\cite{Andres:2022ovj,Andres:2023xwr,Andres:2023ymw,Devereaux:2023vjz}. See also~\cite{Neill:2022lqx} for a review on EEC in the precision QCD.

The EEC has exhibited its perturbative simplicity in the fixed order calculations. Advancements in loop calculations have enabled the calculation of EEC analytically at next-to-leading order~(NLO) in QCD~\cite{Dixon:2018qgp,Luo:2019nig} and numerically at NNLO~\cite{DelDuca:2016ily}. The counterpart in the conformally invariant $\mathcal{N}=4$ super-Yang-Mills (SYM) theory has also been calculated analytically at NLO~\cite{Belitsky:2013ofa,Belitsky:2013bja,Belitsky:2013xxa} and NNLO~\cite{Henn:2019gkr} in the weak coupling limit. In addition, the EEC has been explored in the context of the AdS/CFT correspondence, making its strong coupling limit accessible for $\mathcal{N}=4$ SYM~\cite{Hofman:2008ar}. There are also recent calculations on EEC inside the heavy states both at weak and strong coupling~\cite{Chicherin:2023gxt}.

There are two special end-point regions for EEC -- the back-to-back limit $(z\to 1)$ and the collinear limit $(z\to 0)$. Significant progress has been made in understanding factorization and resummation structures in these limit. In the back-to-back limit, the $\mathrm{N^4LL}$ accuracy prediction is achieved in~\cite{Duhr:2022yyp}, based on the all-order understanding of logarithmic structure in~\cite{Moult:2018jzp}. The NNLL resummation result was matched to NNLO fixed order calculation to measure the strong coupling constant $\alpha_s$~\cite{Kardos:2018kqj}.

In the collinear limit, a factorization formula beyond leading logarithmic accuracy was proposed in~\cite{Dixon:2019uzg}, generalizing the previous result based on jet calculus~\cite{Konishi:1979cb}.
The factorization formula is as follows
\begin{equation}
\frac{\langle \mathcal{E}({n})\mathcal{E}({n}^\prime) \rangle}{Q^2} = \frac{1}{z} \sum_a \int_0^1 dx\, x^2 j_a(\ln \frac{z x^2 Q^2}{\mu^2}; \alpha_s(\mu)) H_a(x, \ln \frac{Q^2}{\mu^2};\alpha_s(\mu))\,, \label{eq:EEC_factorization}
\end{equation}
where $j_a$ are the jet functions corresponding to the differential EEC~\footnote{The jet functions $J_a$ in~\cite{Dixon:2019uzg} correspond to the cumulant EEC, which are the integrated version of $j_a$ in this paper.} and $H_a$ are the associated hard functions.
This factorization formula reflects the intuitive picture for the collinear EEC: hard scattering process produces a parton $a$ with momentum fraction $x$, which later undergoes the fragmentation process and produces the energy flows measured by the detectors along $\vec{n}$ and $\vec{n}^\prime$. The same technique has also been employed to study the collinear limit of higher point projective EECs~\cite{Lee:2022ige,Chen:2023zlx}. The factorization formula shows the universality of the collinear behavior of EEC, which is not sensitive to the details of the hard process and the types of the collider. Therefore, collinear EECs are good jet substructure observables as long as the opening angles are much smaller than the jet size, whose distribution provide a visualization of the parton fragmentation evolution and hadronization~\cite{Komiske:2022enw}. Based on collinear EECs, CMS collaboration has the most precise extraction of $\alpha_s$ using jet substructure observables to date~\cite{CMS:2023wcp}. 

Apart from precision calculations and phenomenological applications, the EEC and its generalizations have become an innovative playground where formal theory techniques and QCD phenomenology intersect. This is particularly the case when exploring the end-point regions. For example, the back-to-back limit of EEC is related to the double lightcone limit of local correlation function, where lightcone bootstrap techniques like large spin perturbation theory are applicable~\cite{Korchemsky:2019nzm,Chen:2023wah}. 

Recently, the light-ray operator formalism has shed light on the collinear limit of EEC in Conformal Field Theory~(CFT). Light-ray operators are non-local operators defined on a null line or a null cone~\cite{Kravchuk:2018htv}, with the energy flow operator $\mathcal{E}(n)$ being the most prominent example. They are intrinsic Lorentzian objects and have revealed nice analytic structures in CFT that are not obvious in Euclidean signature~\cite{Caron-Huot:2017vep,Kravchuk:2018htv,Caron-Huot:2022eqs}. It also provides a new proof for the $a$-theorem from ANEC~\cite{Hartman:2023qdn}. When it comes to collider physics, the collinear limit of celestial observables like EEC is governed by the Operator Product Expansion~(OPE) of light-ray operators~\cite{Hofman:2008ar,Kologlu:2019mfz,Chang:2020qpj}, which further leads to the notion of the celestial blocks~\cite{Kologlu:2019mfz,Chang:2020qpj,Chang:2022ryc,Chen:2022jhb}. Despite the significant success of the light-ray OPE in CFT, it remains unclear how to extend this approach to non-conformal theory like QCD.  

In this paper, we are going to make concrete connection between the light-ray OPE and factorization for collinear EEC in perturbative QCD. More specifically, our aim is to extend the light-ray OPE formalism to non-conformal theories such as QCD and derive the factorization formula \eqref{eq:EEC_factorization}, based on the recent progress on the renormalization of light-ray operators in the weakly-coupled theories~\cite{Caron-Huot:2022eqs}. The outline is as follows. In Section \ref{sec:light_ray_operator}, we review the general properties of light-ray operators along with their corresponding renormalization within perturbation theory. As an example, we use the leading twist light-ray operators in QCD to demonstrate these concepts. In Section \ref{sec:light_ray_OPE}, we propose the leading power light-ray OPE ansatz in perturbative QCD and constrain the Wilson coefficients from the Lorentz symmetry, renormalization group invariance, and physical constraints. In Section \ref{sec:QCD_factorization}, we derive the factorization formula \eqref{eq:EEC_factorization} from the light-ray OPE ansatz. In Section \ref{sec:integral_form}, we present an alternative light-ray OPE ansatz in integral form. In Section \ref{sec:conclusion}, we conclude and discuss future directions.


\section{Review of light-ray operators and renormalization}\label{sec:light_ray_operator}

In this section, we briefly review the renormalization of light-ray operators in perturbation theory. This process is associated with the infrared behavior of detectors at infinity. It has been used phenomenologically to resum the spin correlation in QCD at LL accuracy ~\cite{Chen:2020adz, Chen:2021gdk}. A more systematic and detailed study was undertaken in the inspiring work ~\cite{Caron-Huot:2022eqs}, where the level crossing near the Regge intercept is elegantly explained through the mixing of light-ray operators and their corresponding celestial shadows. We recommend interested readers to refer to the original papers~\cite{Kravchuk:2018htv, Caron-Huot:2022eqs} for detailed discussions presented in the following subsections.

\subsection{Properties of light-ray operators}

The simplest kind of light-ray operators are the light transform of local primary operators, similar to the energy flow operator $\mathcal{E}(n)$:
\begin{equation}\label{eq:light_transform_def}
    \mathbb{O}_{\Delta, J}(n) = \lim_{\bar{n}\cdot x \to\infty} \left(\frac{\bar{n}\cdot x}{2}\right)^{\Delta-J} \int_{-\infty}^\infty d(n\cdot x)\, \mathcal{O}_{\Delta,J}(x;\bar{n}) \,,
\end{equation}
where $n=(1,\vec{n})$ and $\bar{n}=(1,-\vec{n})$ are two null vectors.
$\mathcal{O}_{\Delta,J}(x;z)$ is the index-free notation for a local primary operator with definite scaling dimension $\Delta$ and spin $J$
\begin{equation}
    \mathcal{O}_{\Delta,J}(x;z) = \mathcal{O}_{\Delta,J}^{\mu_1\cdots\mu_J}(x) z_{\mu_1}\cdots z_{\mu_J} \,.
\end{equation}
From the definition \eqref{eq:light_transform_def}, we can see that the light-ray operators has scaling dimension $J-1$~\footnote{In~\cite{Kravchuk:2018htv}, the light transform of $\mathcal{O}_{\Delta,J}$ has the dimension $1-J$. The difference is due the fact that the detectors are inserted at spatial infinity, which has the opposite dimension compared with the corresponding operator located at a generic point.} and collinear spin $1-\Delta$ with respect to the boost generator $\vec{\mathbf{K}}$ along $\vec{n}$
\begin{equation}
    i[\vec{n}\cdot \vec{\mathbf{K}},\mathbb{O}_{\Delta,J}(n)] = -(1-\Delta) \mathbb{O}_{\Delta,J}(n) \,.
\end{equation}
Therefore, the light transform somehow switches the role of the dimension and the collinear spin in conformal theory.
By imposing the scaling property
\begin{equation}
    \mathbb{O}_{\Delta,J}(\lambda n) = \lambda^{1-\Delta} \mathbb{O}_{\Delta,J}(n) \,,
\end{equation}
the light-ray operators will transform covariantly under the Lorentz group action
\begin{equation}
    U_\Lambda \mathbb{O}_{\Delta,J}(n) U_\Lambda^\dagger = \mathbb{O}_{\Delta,J}(\Lambda n) \,.
\end{equation}

In free theories,  the meaning of light-ray operators becomes clear for the twist-2 operators. Taking massless QCD for example, we have the following local composite operators without transverse spin
\begin{align}
    &\mathcal{O}_{q}^{[J]}(x;\bar{n})=\frac{1}{2^J}\bar{\psi}(x) \slashed{\bar{n}} (i \bar{n}\cdot D)^{J-1} \psi(x)\,,\\
    &\mathcal{O}_{g}^{[J]}(x;\bar{n})=\frac{1}{2^J}\bar{n}_\mu\bar{n}_{\nu} F_c^{\mu\rho}(x)(i \bar{n}\cdot D)^{J-2} F_c^{\nu\rho}(x)\,,
\end{align}
where $D_\mu$ is the covariant derivative and $F_c^{\mu\nu}$ is the gluon field strength tensor with adjoint color index $c$. In the free case, we can use the mode expansions
\begin{align}
    \psi(x) &= \sum_{s=\pm} \int \frac{d^3\vec{p}}{(2\pi)^3 2E}  \left[ b_{p,s} u_s(p) e^{-i p\cdot x} + d_{p,s}^\dagger v_s(p) e^{i p\cdot x} \right] \,,\\
    A_c^\mu(x) &= \sum_{\lambda =\pm} \int \frac{d^3\vec{p}}{(2\pi)^3 2E}  \left[ a_{p,\lambda,c} \epsilon_{\lambda}^\mu(p) e^{-i p\cdot x} + a_{p,\lambda,c}^\dagger \epsilon_{\lambda}^{*\mu}(p) e^{i p\cdot x} \right] \,,
\end{align}
the light-ray operators are then given by, after normal ordering,
\begin{align}
    \mathbb{O}_q^{[J]}(n) &= \sum_{s} \int \frac{E^2 dE}{(2\pi)^3 2E} E^{J-1} \left(b_{p,s}^\dagger b_{p,s} + (-1)^J d^\dagger_{p,s} d_{p,s} \right)\,,\\
    \mathbb{O}_g^{[J]}(n) &= \frac{1+(-1)^J}{2}\sum_{\lambda, c} \int \frac{E^2 dE}{(2\pi)^3 2E} E^{J-1} a_{p,\lambda, c}^\dagger a_{p,\lambda, c} \,.
\end{align}
We can gain a glimpse of the meaning and properties of light-ray operators from these expressions. First, in the free theory, $\mathbb{O}_q^{[J]}(n)$ and $\mathbb{O}_g^{[J]}(n)$ respectively measure the quark and gluon flowing along direction $\vec{n}$ and weight them with $E^{J-1}$, which also reflects the general property of light transform on switching the role of dimension and spin. Second, while local operators are only well-defined at integer spin $J$, light-ray operators can be analytically continued in spin, realizing the concept of Regge trajectory in CFT~\cite{Caron-Huot:2017vep,Kravchuk:2018htv}. After analytic continuation, the light-ray operators are no longer the light transform of local operators. Third, the presence of $(-1)^J$ signals the fact that the analytic continuations are distinct for odd and even spin respectively, which is related to the CRT transformation properties of light-ray operators~\cite{Kravchuk:2018htv}. In the remainder of this paper, as the odd spin branch is irrelevant to the EEC, we will focus solely on the even spin branch and replace $(-1)^J$ by $1$ in the above expressions. One method to detect the odd spin branch involves considering the EEC between like-sign charged particles~\cite{Lee:2023xzv,Lee:2023tkr}, which is not an IRC observable and is beyond the scope of this paper.

\subsection{Renormalization of light-ray operators}

The discussion in the previous subsection is valid for CFT or massless free theory. In an interacting theory, light-ray operators have infrared divergences and require renormalization to be well-defined. In particular, the dimension is no longer a good quantum number in non-conformal theory. In the following, we will take the twist-2 operators in massless QCD as an example to illustrate the renormalization of light-ray operators.

The starting point is the bare light-ray operators $\mathbb{O}_{a;\text{bare}}^{[J]}(n)$, which is defined by the light transform of the bare local operators $\mathcal{O}_{a;\text{bare}}^{[J]}(x;\bar{n})$:
\begin{equation}
    \mathbb{O}_{a;\text{bare}}^{[J]}(n) = \lim_{\bar{n}\cdot x \to\infty} \left(\frac{\bar{n}\cdot x}{2}\right)^{2} \int_{-\infty}^\infty d(n\cdot x)\, \mathcal{O}_{a;\text{bare}}^{[J]}(x;\bar{n}) \,,
\end{equation}
where we have set the bare dimension of the local operator to be $\Delta^{\text{bare}}=J+d-2=J+2$ in $d=4$ spacetime dimension.
Here, the index $a=q,g$ denotes the quark and gluon operators respectively. 
The bare light-ray operators $\mathbb{O}_{a;\text{bare}}^{[J]}$ have dimension $\Delta_L^{\text{bare}}=J-1$ and Lorentz spin $J_L^{\text{bare}}=1-\Delta^{\text{bare}}=-1-J$ where the subscript $L$ stands for light-ray operators. Inside correlation functions, the bare operators are often IR divergent after including loop corrections, and hence need to be renormalized. 

We employ the dimensional regularization and $\overline{\text{MS}}$-scheme for the renormalization. One important property is that operators with different symmetry cannot mix with each other. In terms of light-ray operators, as pointed out in~\cite{Caron-Huot:2022eqs}, renormalization does not change the Lorentz spin $J_L^{\text{ren}}=J_L^{\text{bare}}\equiv J_L$ and operators with different spins cannot mix. Moreover, in dimensional regularization where $d=4-2\epsilon$, operators with different classical/bare dimensions cannot mix either, except when the difference is of order $\epsilon$. Therefore, the renormalization of twist-2 light-ray operators, $\mathbb{O}_{q}^{[J]}$ and $\mathbb{O}_g^{[J]}$, is particularly simple -- they can only mix with each other when they have the same label $J$. At the same time, $J$ should be large enough to be away DGLAP/BFKL mixing~\cite{Brower:2006ea,Jaroszewicz:1982gr,Lipatov:1996ts,Kotikov:2000pm,Kotikov:2002ab,Kotikov:2007cy}, which is satisfied for the light-ray OPE in EEC~\cite{Kologlu:2019mfz,Kologlu:2019bco}.

The renormalization procedure is standard. We introduce the renormalized light-ray operators $\mathbb{O}_{a;\text{ren}}^{[J]}$, which are related to the bare operators through the renormalization matrix $\mathbb{Z}_{ab}^{[J]}$:
\begin{equation}
    \mathbb{O}^{[J]}_{a;\text{bare}} = \mathbb{Z}^{[J]}_{ab} \mathbb{O}^{[J]}_{b;\text{ren}} \,.
\end{equation}
Note that for the renormalized operator $\mathbb{O}_{a;\text{ren}}^{[J]}$, 
the label $J$ does not have the interpretation of spin any more. Instead, a more appropriate label is the Lorentz spin of the light-ray operators $J_L$: $\mathbb{O}_{a;\text{ren}}^{\{J_L\}}\equiv \mathbb{O}_{a;\text{ren}}^{[J=-1-J_L]}$ when $d=4$. But nevertheless, we will still use the label $J$ in order to compare with the QCD factorization formula in the Section \ref{sec:QCD_factorization}.

The renormalized operators $\mathbb{O}_{a;\text{ren}}^{[J]}$ have scale dependence, which is governed by the renormalization group equation (RGE):
\begin{equation}
    \frac{d}{d\ln\mu^2} \mathbb{O}_{a;\text{ren}}^{[J]}(n;\mu) = \gamma_{ab}^T(J;\alpha_s(\mu)) \mathbb{O}_{b;\text{ren}}^{[J]}(n;\mu) \,,\label{eq:op_RGE}
\end{equation}
where the anomalous dimension matrix $\gamma_{ab}^T(J;\alpha_s(\mu))$ is a function of the coupling constant $\alpha_s(\mu)$ and the label $J$, which can be extracted from the $1/\epsilon$ pole of the renormalization factor $\mathbb{Z}_{ab}^{[J]}$. Later, we will see that the anomalous dimension matrix $\gamma_{ab}^T(J;\alpha_s(\mu))$ is closely related to time-like splitting function, which is the kernel of the DGLAP evolution equation~\cite{Altarelli:1977zs,Dokshitzer:1977sg,Gribov:1972ri} in fragmentation functions~\cite{deFlorian:2014xna,Anderle:2015lqa,Chen:2020uvt}.

As shown in~\cite{Caron-Huot:2022eqs}, the light-ray operator renormalization picture provides an elegant CFT explanation to the reciprocity relation~\cite{Gribov:1972rt,Basso:2006nk,Dokshitzer:2006nm,Dokshitzer:2005bf,Marchesini:2006ax,Mueller:1982cq,Neill:2020bwv}. This relation connects the time-like anomalous dimension to the space-like counterpart, if there is no mixing:
\begin{align}
    2\gamma^S(J;\alpha_s) = 2\gamma^T(J+2\gamma^S(J;\alpha_s);\alpha_s)\,,\label{eq:reciprocity_relation}\\
    2\gamma^T(J;\alpha_s) = 2\gamma^S(J-2\gamma^T(J;\alpha_s);\alpha_s)\,.
\end{align}
When quark-gluon mixing is present, it has been checked that reciprocity relation still holds, in the sense of eigenvalues, to the three loop in QCD~\cite{Chen:2020uvt}. The reciprocity relation is important for the fact that the collinear limit of EEC in CFT is governed by the space-like anomalous dimension, though the underlying processes are time-like splittings~\cite{Dixon:2019uzg}.


\section{Light-ray OPE in perturbation theory}\label{sec:light_ray_OPE}

Light-ray operators, such as \eqref{eq:light_transform_def}, behave like local operators on the celestial sphere. The collinear limit in collider physics corresponds to the coincident limit on the celestial sphere for the celestial observables like EEC, motivating the notion of OPE for light-ray operators. This idea is originated from the novel work of Hofman and Maldacena~\cite{Hofman:2008ar} on conformal collider physics, and was systematically developed as the light-ray OPE in CFT~\cite{Kologlu:2019mfz,Chang:2020qpj}.

The general form of the light-ray OPE in CFT is determined by symmetry. Suppose we have two light-ray operators $\mathbb{O}_{\Delta_1,J_1}(n)$ and $\mathbb{O}_{\Delta_2,J_2}(n^\prime)$, their OPE has the schematic form
\begin{equation}
    \mathbb{O}_{\Delta_1,J_1}(n) \mathbb{O}_{\Delta_2,J_2}(n^\prime) \supset \theta^{\Delta-\Delta_1-\Delta_2+1} \mathbb{O}_{\Delta,J}(n)\,,\quad \theta \to 0 \,,\label{eq:light_ray_OPE_schematic}
\end{equation}
where $n\cdot n^\prime=1-\cos\theta$. By counting the scaling dimension on both sides, we find the constraint on the spin $J=J_1+J_2-1$~\cite{Kologlu:2019mfz,Kologlu:2019bco}. The exponent on $\theta$ is determined by the boost symmetry along $\vec{n}$. The systematic expansion is organized by the celestial blocks~\cite{Kologlu:2019mfz,Chang:2020qpj}, which is not necessary at the leading power.

In non-conformal theories, scaling symmetry is broken and the constraint $J=J_1+J_2-1$ is no longer valid. However, we would expect the violation is small in perturbative QCD because QCD is classically conformal in $4d$ and the breaking is caused by the running of the coupling constant. 

In perturbative QCD, we consider EEC $\langle \mathcal{E}(n)\mathcal{E}(n^\prime)\rangle$ for simplicity because the energy flow operator $\mathcal{E}(n)$, as the light transform of the stress tensor, is a twist-2 light-ray operator with vanishing anomalous dimension. In conformal theories, we expect only $J=3$ light-ray operators $\mathbb{O}_{\Delta,J=3}$ can appear in the $\mathcal{E}\mathcal{E}$ OPE. To include scaling symmetry breaking, we consider the following ansatz for the light-ray OPE
\begin{equation}
    \mathcal{E}(n)\mathcal{E}(n^\prime) = \sum_{k=0}^\infty C_a^{(k)}(z;\mu) \left[\partial_J^k \mathbb{O}^{[J]}_{a;\text{ren}}(n;\mu)\right]\bigg|_{J=3} + \text{higher twists}\,, \label{eq:light_ray_OPE_ansatz}
\end{equation}
where $z=n\cdot n^\prime/2$ and $\partial_J$ is the derivative acting on the label $J$ (or equivalently on the Lorentz spin $J_L$). 
Here, we assume that the light-ray operators $\mathbb{O}^{[J]}_{a;\text{ren}}(n;\mu)$ are analytic with respect to $J$ to make sense of the derivative actions, which we believe to be plausible in perturbation theory, at least up to any finite loop orders.
The higher twist terms are suppressed by powers of $z$ and will be neglected in the following. The coefficients $C_a^{(k)}(z;\mu)$ are the Wilson coefficients, which are functions of $z$ and the renormalization scale $\mu$ (including both $\alpha_s(\mu)$ and explicit $\mu$ dependence). 

The evaluation point is chosen to be $J=3$ for convenience because this is the case for classical conformal theory. As long as the analyticity holds, we can choose any (nearby) evaluation point in principle and Taylor expansion relates them by a change of basis
\begin{equation}
    \left[\partial_J^k \mathbb{O}^{[J]}_{a,\text{ren}}(\vec{n};\mu)\right]\Bigg|_{J=3+\delta} = \left[\partial_J^k \mathbb{O}^{[J+\delta]}_{a,\text{ren}}(\vec{n};\mu)\right]\Bigg|_{J=3} =\sum_{K=0}^{\infty}\frac{\delta^K}{K!} \left[\partial_J^{k+K} \mathbb{O}^{[J]}_{a,\text{ren}}(\vec{n};\mu)\right]\Bigg|_{J=3}\,.
\end{equation}

In the following subsections, we will constrain the Wilson coefficients $C_a^{(k)}(z;\mu)$ from the Lorentz symmetry, RG invariance, and observables, which is important to relate to the QCD factorization formula in the Section \ref{sec:QCD_factorization}.

\subsection{Lorentz symmetry}

As we have seen in \eqref{eq:light_ray_OPE_schematic}, the power of $\theta$ is determined by the boost symmetry along $\vec{n}$ in CFT. This is suffice for the leading power consideration and the full Lorentz symmetry lead to the notion of celestial blocks~\cite{Kologlu:2019mfz,Chang:2020qpj}. Therefore, in this subsection, we only explore the implication of the boost symmetry along $\vec{n}$ for the ansatz \eqref{eq:light_ray_OPE_ansatz}.

For concreteness, we specify a frame for the null vectors $n,n^\prime$ using spherical coordinates
\begin{equation}
n = (1,0,0,1),\,\quad n^\prime(\theta,\phi) = (1, \sin\theta \cos\phi,\sin\theta \sin\phi,\cos\theta)\,.
\end{equation}
Now we apply the boost generator $\vec{n}\cdot \vec{\mathbf{K}}$ to the operators on both sides the light-ray OPE ansatz \eqref{eq:light_ray_OPE_ansatz}
\begin{align}
    &\text{L.H.S. : } -i[\vec{n}\cdot \vec{\mathbf{K}}, \mathcal{E}({n})\mathcal{E}({n}^\prime(\theta,\phi))] = (-3(1+\cos\theta)-\sin\theta \partial_\theta) \mathcal{E}({n})\mathcal{E}({n}^\prime(\theta,\phi)) \,,\\
    &\text{R.H.S. : } -i[\vec{n}\cdot \vec{\mathbf{K}}, \partial_J^k \mathbb{O}^{[J]}_{a;\text{ren}}({n};\mu)]=-(J+1)\partial^k_J \mathbb{O}^{[J]}_{a;\text{ren}}({n};\mu) -k \partial_J^{k-1} \mathbb{O}^{[J]}_{a;\text{ren}}({n};\mu)\,,
\end{align}
where the second line is from the fact that $\mathbb{O}^{[J]}_{a;\text{ren}}({n};\mu)$ is a light-ray operator with Lorentz spin $J_L=-1-J$. In the collinear limit $\theta\to 0$, we can approximate the differential operator in the first line by a homogeneous differential operator in $z$
\begin{equation}
    (-3(1+\cos\theta)-\sin\theta \partial_\theta)\approx -2(3+z\partial_z)\,.
\end{equation}
and hence establish the equality
\begin{equation}
    \begin{split}
        &-2(3+z\partial_z) \sum_{k=0}^\infty C_a^{(k)}(z;\mu) \left[\partial_J^k \mathbb{O}_{a;\text{ren}}^{[J]}(\vec{n};\mu)\right]\Bigg|_{J=3}\\
        =& \sum_{k=0}^\infty C_a^{(k)}(z;\mu) \left[-(J+1)\partial^k_J \mathbb{O}^{[J]}_{a;\text{ren}}(\vec{n};\mu) -k \partial_J^{k-1} \mathbb{O}^{[J]}_{a;\text{ren}}(\vec{n};\mu)\right]\Bigg|_{J=3}\,.
    \end{split}
\end{equation}
This leads to the recursion relation on the Wilson coefficients $C_a^{(k)}(z;\mu)$:
\begin{equation}
\left(\frac{\partial}{\partial \ln z} + 1\right) C_a^{(k)}(z;\mu) = \frac{k+1}{2} C_a^{(k+1)}(z;\mu)\,.
\end{equation}
The solution can be determined up to an arbitrary function $\widetilde{C}_a(z;\mu)$
\begin{equation}
\boxed{
C_a^{(k)}(z;\mu) =\frac{1}{z} \frac{2^k}{k!} \left(\frac{\partial}{\partial \ln z}\right)^k \widetilde{C}_a(z;\mu)
}
\,.\label{eq:Ck_from_Lorentz}
\end{equation}

Therefore, although we introduce infinitely many Wilson coefficients $C_a^{(k)}(z;\mu)$ in the ansatz \eqref{eq:light_ray_OPE_ansatz}, only the first one is arbitrary and the rest can be determined by the Lorentz symmetry. This is consistent with the picture in CFT that only a few OPE coefficients show up at the leading power. We define the function $\widetilde{C}_a(z;\mu)$ in a way that does not contain the classical scaling behavior $\sim 1/z$ and hence we expect that, at any finite order in perturbation theory, it contains only logarithms in $z$.

\subsection{RG invariance}

The left-hand side of the light-ray OPE ansatz \eqref{eq:light_ray_OPE_ansatz} is RG invariant due to the vanishing anomalous dimension of $\mathcal{E}$:
\begin{equation}
    \frac{d}{d\ln\mu^2} \left[\mathcal{E}(n)\mathcal{E}(n^\prime)\right] = 0 \,,
\end{equation}
which imposes the RG consistency conditions on the right-hand side. The RG equation for the operator $\partial_J^k \mathbb{O}^{[J]}_{a;\text{ren}}(\vec{n};\mu)$ follows straightforwardly from \eqref{eq:op_RGE}:
\begin{equation}
    \frac{d}{d\ln \mu^2} \left[\partial_J^k \mathbb{O}^{[J]}_{a;\text{ren}}({n};\mu)\right] = \sum_{p=0}^{k} \frac{k!}{p!(k-p)!} \left(\partial_J^{k-p} \gamma^T_{ab}(J;\alpha_s(\mu) )\right) \partial_J^p \mathbb{O}^{[J]}_{a;\text{ren}}({n};\mu)\,.
\end{equation}
The RG invariance leads to the evolution equation for the Wilson coefficients $C_a^{(k)}(z;\mu)$
\begin{equation}
    \frac{d}{d\ln \mu^2} C_a^{(k)}(z;\mu) = -\sum_{m=0}^{\infty} \frac{(k+m)!}{k!\, m!} C_b^{(k+m)}(z;\mu) \partial_J^m \gamma^T_{ba}(J;\alpha_s(\mu))\,.
\end{equation}
Combined with (\ref{eq:Ck_from_Lorentz}), we find that infinitely many evolution equations reduce to single one on $\widetilde{C}_a(z;\mu)$:
\begin{equation}
\boxed{
\frac{d}{d\ln \mu^2} \widetilde{C}_a(z;\mu) = -\sum_{m=0}^\infty \frac{2^m}{m!} \left(\frac{\partial}{\partial \ln z}\right)^m \widetilde{C}_b(z;\mu) \left[\partial_J^m \gamma^T_{ba}(J;\alpha_s(\mu))\right]\bigg|_{J=3}
}\,, \label{eq:RG_C}
\end{equation}
showing the compatibility between the constraints from RG invariance and Lorentz symmetry. In Section \ref{sec:QCD_factorization}, we will see that this evolution equation is equivalent to the RG equation of jet functions.

\subsection{Constraints from observables}

Up to now, the undetermined coefficient $\widetilde{C}_a(z;\mu)$ is an arbitrary multi-variable function, unlike the simplicity of its counterpart in CFT. In this subsection, we will show that by considering the physical observable EEC, we can further constrain the function $\widetilde{C}_a(z;\mu)$ down to a single-variable function with respect to a particular combination of $z$ and $\mu$, if we neglect the implicit $\mu$-dependence in the coupling $\alpha_s(\mu)$.

We consider the EEC with the center of mass energy $Q$. The physical input is that, in perturbative massless QCD, EEC has the following functional form
\begin{equation}
    \langle\mathcal{E}(n)\mathcal{E}(n^\prime)\rangle_Q = Q^2 f(z,\ln \frac{Q^2}{\mu^2},\alpha_s(\mu))\,, \label{eq:EEC_functional_form}
\end{equation}
where $f$ is a dimensionless function that depends on energy scales either through ratios or implicitly through the running coupling of $\alpha_s$. After applying the light-ray OPE ansatz \eqref{eq:light_ray_OPE_ansatz}, we find
\begin{equation}
    \langle\mathcal{E}(n)\mathcal{E}(n^\prime)\rangle_Q = \frac{1}{z} \sum_{k=0}^{\infty} \frac{2^k}{k!} \left[\left(\frac{\partial}{\partial \ln z}\right)^k \widetilde{C}_a(z;\mu)\right] \left[\partial_J^k \langle \mathbb{O}^{[J]}_{a;\text{ren}}(n;\mu)\rangle_Q\right]\Big|_{J=3}\,.
\end{equation}
In perturbation theory, the bare operator $\mathbb{O}^{[J]}_{a;\text{bare}}$ measures $E^{J-1}$ when $d=4$, so we expect the one-point correlator $\langle \mathbb{O}^{[J]}_{a;\text{ren}}(n;\mu)\rangle_Q$ takes the form
\begin{equation}
\langle \mathbb{O}^{[J]}_{a;\text{ren}}(n;\mu)\rangle_Q = Q^{J-1} h_a(J,\ln\frac{Q^2}{\mu^2};\alpha_s(\mu))\,.
\end{equation}
Later, we will see $h_a$ is the moment of the hard function $H_a$ in the factorization formula. 

At the first sight, the light-ray OPE formalism seem to screw up our desired functional form \eqref{eq:EEC_functional_form} of EEC:
\begin{equation}
    \begin{split}
    &\langle\mathcal{E}(n)\mathcal{E}(n^\prime)\rangle_Q \\
     =& \frac{Q^2}{z} \sum_{k=0}^\infty \frac{2^k}{k!}\left[\left(\frac{\partial}{\partial \ln z}\right)^k \widetilde{C}_a(z;\mu)\right] 
\sum_{m=0}^{k} \frac{k!}{m!(k-m)!} (\ln Q)^{k-m} \left[\partial_J^m h_a(J,\ln\frac{Q^2}{\mu^2};\alpha_s(\mu))\right]\Bigg|_{J=3}.
    \end{split}
\end{equation}
However, we can rearrange the summation order and obtain
\begin{equation}
    \langle\mathcal{E}(n)\mathcal{E}(n^\prime)\rangle_Q
    = \frac{Q^2}{z} \sum_{m=0}^\infty \frac{1}{m!} \left[\partial_J^m h_a(J,\ln\frac{Q^2}{\mu^2};\alpha_s(\mu))\right]\Bigg|_{J=3} \times 2^m \left(\frac{\partial}{\partial \ln z}\right)^m \widetilde{C}_a(zQ^2;\mu)\,, \label{eq:EEC_functional_form_from_OPE}
\end{equation}
where we have reversely used Taylor expansion at the same time
\begin{equation}
    \sum_{k=m}^\infty \frac{ (2 \ln Q)^{k-m}}{(k-m)!}  \left(\frac{\partial}{\partial \ln z}\right)^{k-m} \widetilde{C}_a(z;\mu)
= \sum_{k=0}^\infty \frac{(\ln Q^2)^k}{k!} \left(\frac{\partial}{\partial \ln z}\right)^k \widetilde{C}_a(z;\mu) = \widetilde{C}_a(zQ^2;\mu)\,.
\end{equation}
In order to match the functional form \eqref{eq:EEC_functional_form} where the explicit $\mu$ dependence only shows up as the combination of the ratio $\frac{Q}{\mu}$, we have to impose:
\begin{equation}
\boxed{
\widetilde{C}_a(zQ^2;\mu) = \widetilde{C}_a(\ln \frac{zQ^2}{\mu^2};\alpha_s(\mu))\,,
\qquad \text{or} \qquad \widetilde{C}_a(z;\mu) = \widetilde{C}_a(\ln \frac{z}{\mu^2};\alpha_s(\mu))
}\,. \label{eq:constraint_on_Ctilde}
\end{equation}
Therefore, we have shown that the explicit $\mu$ dependence in the Wilson coefficient $\widetilde{C}_a(z;\mu)$ is only through the combination $\ln \frac{z}{\mu^2}$, which is the same as the jet function $j_a$ in the QCD factorization formula \eqref{eq:EEC_factorization}.


\section{Relation to QCD factorization formula}\label{sec:QCD_factorization}

In this section, we are ready to relate the light-ray OPE formalism to the QCD factorization formula for EEC \eqref{eq:EEC_factorization} based on the various constraints on the light-ray OPE ansatz \eqref{eq:light_ray_OPE_ansatz} from the previous section.

We express $h_a(J)$ as the Mellin moment of a function $\widetilde{h}_a(x)$
\begin{equation}
h_a(J,\ln\frac{Q^2}{\mu^2};\alpha_s(\mu)) = \int_0^1 dx\, x^{J-1} \widetilde{h}_a(x,\ln \frac{Q^2}{\mu^2};\alpha_s(\mu))\,,
\end{equation}
and apply it to the EEC \eqref{eq:EEC_functional_form_from_OPE}
\begin{align}
    \frac{\langle\mathcal{E}(n)\mathcal{E}(n^\prime)\rangle_Q}{Q^2}
    =& \frac{1}{z}\int_0^1 dx\, x^2 \widetilde{h}_a(x,\ln \frac{Q^2}{\mu^2};\alpha_s(\mu)) \sum_{m=0}^\infty \frac{2^m}{m!} (\ln x)^m \left(\frac{\partial}{\partial \ln z}\right)^m \widetilde{C}_a(\ln\frac{z Q^2}{\mu^2}; \alpha_s(\mu))\nonumber\\
    =& \frac{1}{z}\int_0^1 dx\, x^2 \widetilde{h}_a(x,\ln \frac{Q^2}{\mu^2};\alpha_s(\mu)) \widetilde{C}_a(\ln\frac{z x^2 Q^2}{\mu^2}; \alpha_s(\mu))\,,
\end{align}
where we use the same trick as in the previous section to switch the order of summation and integration, followed by reversely using Taylor expansion. The last line is the QCD factorization formula \eqref{eq:EEC_factorization} with $\widetilde{C}_a$ and $\widetilde{h}_a$ being identified as the jet function $j_a$ and hard function $H_a$ respectively. This is the main result of this paper, which is in perfect agreement with the factorization formula in~\cite{Dixon:2019uzg} after integrating over $z$.

We can re-write the RG equation for $\widetilde{C}_a$ \eqref{eq:RG_C} in a similar way
\begin{align}
&\frac{d}{d\ln \mu^2} \widetilde{C}_a(\ln\frac{z  Q^2}{\mu^2}; \alpha_s(\mu)) \nonumber\\
=& \int_0^1 dy\, y^2 P^T_{ba}(y;\alpha_s(\mu)) \sum_{m=0}^\infty \frac{2^m}{m!} (\ln y)^m \left(\frac{\partial}{\partial \ln z}\right)^m \widetilde{C}_b(\ln\frac{z Q^2}{\mu^2}; \alpha_s(\mu))\nonumber\\
=& \int_0^1 dy\, y^2 \widetilde{C}_b(\ln\frac{z y^2 Q^2}{\mu^2}; \alpha_s(\mu)) P^T_{ba}(y;\alpha_s(\mu))\,.
\end{align}
Here we have used the relation that the anomalous dimension matrix $\gamma^T_{ba}(J;\alpha_s(\mu))$ is the Mellin transform of the time-like splitting function $P^T_{ba}(y;\alpha_s(\mu))$:
\begin{equation}
    \gamma_{ab}^T(J;\alpha_s(\mu)) = - \int_0^1 dy\, y^{J-1} P^T_{ab}(y;\alpha_s(\mu))\,.
\end{equation}
Therefore, the RG equation for the Wilson coefficient \eqref{eq:RG_C} reproduce the EEC jet function evolution in~\cite{Dixon:2019uzg}.

%
%
%
%
%
%

\section{Integral form of light-ray OPE ansatz}\label{sec:integral_form}

In this section, we propose an alternative integral form of the light-ray OPE ansatz \eqref{eq:light_ray_OPE_ansatz}. This new form is not only more adaptable to general cases but also makes the derivation in the preceding sections more transparent. The idea is to assume all light-ray operators, which are not forbidden by symmetry, can appear in the light-ray OPE. More concretely, the leading twist integral ansatz is~\footnote{We thank David Simmons-Duffin for suggesting this form of light-ray OPE.}
\begin{equation}
    \mathcal{E}(n_1)\mathcal{E}(n_2)=\int dJ_L\, c(J_L;\mu) B_{J_L}(n_1,n_2;\partial_{n_2}) \mathbb{O}^{\{J_L\}}_{\text{ren}}(n_2;\mu) +\text{higher twists}\,, \label{eq:light_ray_OPE_integral_ansatz}
\end{equation}
where the integral is over Lorentz spin $J_L$ and we have dropped the label $a$ for simplicity. $B_{J_L}(n_1,n_2;\partial_{n_2})$ is a differential operator that is completely fixed by Lorentz symmetry~\cite{Kologlu:2019mfz}:
\begin{equation}
    B_{J_L}(n_1,n_2;\partial_{n_2}) = (\frac{n_1\cdot n_2}{2})^{1-d-\frac{J_L}{2}} + \text{higher power}\,.
\end{equation}

To make connection with the differential form \eqref{eq:light_ray_OPE_ansatz}, we choose a reference point for Lorentz spin $J_L=J_L^*+\nu$, where $J_L^*$ is $-4$ in \eqref{eq:light_ray_OPE_ansatz} for spacetime dimension $d=4$. By assuming analyticity in the light-ray operators, we can Taylor expand around $J_L^*$
\begin{align}
    \mathcal{E}(n_1)\mathcal{E}(n_2)&\approx\int d\nu\, c(J_L^*+\nu;\mu) z^{-3-\frac{J_L^*}{2}-\frac{\nu}{2}} \exp(\nu\, \partial_{J_L^\prime})\mathbb{O}^{\{J_L^\prime\}}_{\text{ren}}(n_2;\mu)\bigg|_{J_L^\prime = J_L^*} \,,\\
    &= \frac{1}{z}\exp\left(-2\frac{\partial}{\partial \ln z} \partial_{J_L^\prime}\right)\int d\nu\, z^{-\nu/2} c(\nu-4;\mu)  \mathbb{O}^{\{J_L^\prime\}}_{\text{ren}}(n_2;\mu)\bigg|_{J_L^\prime = -4} \,,
\end{align}
which is consistent with the differential ansatz \eqref{eq:light_ray_OPE_ansatz} and the constraints \eqref{eq:Ck_from_Lorentz} from Lorentz symmetry if we identify
\begin{equation}
    \widetilde{C}(z;\mu) = \int d\nu\, z^{-\nu/2} c(\nu-4;\mu)\,. \label{eq:C_from_c}
\end{equation}
For perturbative QCD where $\widetilde{C}$ only consists of logarithms in $z$ in weak coupling expansion, we expect the $\nu$-integral is localized around $\nu=0$. 

The constraint \eqref{eq:EEC_functional_form} from the energy correlator is also straightforward to impose in the integral form. In $d=4$, the one-point event shape $\langle \mathbb{O}^{\{J_L\}}_{\text{ren}} \rangle$ has the following functional form in perturbation theory
\begin{equation}
    \langle \mathbb{O}^{\{J_L\}}_{\text{ren}}(n;\mu) \rangle_Q = Q^{-J_L-2} h(-J_L-1,\ln \frac{Q^2}{\mu^2};\alpha_s(\mu))\,,
\end{equation}
which leads to collinear limit of EEC as
\begin{equation}
    \langle\mathcal{E}(n)\mathcal{E}(n^\prime)\rangle_Q = \frac{Q^2}{z} \int d\nu\, z^{-\nu/2} c(\nu-4;\mu) Q^{-\nu} h(3-\nu,\ln \frac{Q^2}{\mu^2};\alpha_s(\mu))\,.
\end{equation}
To match the functional form \eqref{eq:EEC_functional_form} in perturbative QCD, we have to impose
\begin{equation}
    c(\nu-4;\mu) = \mu^\nu \tilde{c}(\nu;\alpha_s(\mu))\,.
\end{equation}
This condition can be translated to the constraint \eqref{eq:constraint_on_Ctilde} on $\widetilde{C}(z;\mu)$ by using \eqref{eq:C_from_c}.

Finally, we can derive the RG equation for $\widetilde{C}(z;\mu)$ from the RG equation of $c(J_L;\mu)$. The RG equation for the light-ray operator $\mathbb{O}^{\{J_L\}}_{\text{ren}}(n;\mu)$ is
\begin{equation}
    \frac{d}{d\ln \mu^2} \mathbb{O}^{\{J_L=\nu-4\}}_{\text{ren}}(n;\mu) = \gamma^T(3-\nu;\alpha_s(\mu)) \mathbb{O}^{\{J_L=\nu-4\}}_{\text{ren}}(n;\mu)\,.
\end{equation}
The RG consistency requires that the Wilson coefficient $c(\nu-4;\mu)$ satisfies
\begin{equation}
    \frac{d}{d\ln \mu^2} c(\nu-4;\mu) = -c(\nu-4;\mu) \gamma^T(3-\nu;\alpha_s(\mu))\,,
\end{equation}
which yields the partial differential equation on $\tilde{c}(\nu-4;\alpha_s)$:
\begin{equation}
    \beta(\alpha_s)\frac{\partial}{\partial \alpha_s} \tilde{c}(\nu;\alpha_s) = - \tilde{c}(\nu;\alpha_s) (\nu + 2\gamma^T(3-\nu;\alpha_s))\,.
\end{equation}
These RG equations are Mellin transformation of the RG equation \eqref{eq:RG_C}. 
In a conformal theory, when the beta function $\beta(\alpha_s)$ vanishes, the solution is simply
\begin{equation}
    \tilde{c}(\nu;\alpha_s) \propto \delta(\nu + 2\gamma^T(3-\nu;\alpha_s))\propto \delta(\nu+2\gamma^S(3;\alpha_s))\,,
\end{equation}
where the appearance of the space-like anomalous dimension is due to the reciprocity relation \eqref{eq:reciprocity_relation}. This is the observation made in \cite{Dixon:2019uzg,Kologlu:2019mfz,Korchemsky:2019nzm} that the space-like anomalous dimension $\gamma^S(3)$ governs the scaling behavior of EEC in CFT, manifested in the form $z^{-1+\gamma^S(3)}$.


\section{Discussion}\label{sec:conclusion}

In this paper we have modified the light-ray OPE in CFT to non-conformal theories, enabling the derivation of the QCD factorization formula for EEC~\cite{Dixon:2019uzg} from the light-ray OPE. The form of light-ray OPE is constrained by the Lorentz symmetry, RG invariance, and functional form of EEC in massless perturbative QCD. Our proposed forms of light-ray OPE, at the leading power, is based on the analyticity assumption on the light-ray operators, which is interesting to explore in the future. The convergence of the sum in \eqref{eq:light_ray_OPE_ansatz} or the integral in \eqref{eq:light_ray_OPE_integral_ansatz} is also worth further investigation. Apart from these subtle issues, we believe that the light-ray OPE formalism is a powerful tool to study in perturbation theories and general Lorentzian QFTs. There are a number of future directions that are interesting from both theoretical and phenomenological perspectives. First, it is important to encompass mass effects in the light-ray operator formalism for much broader applications. There are some studies on the effects of quark masses in energy correlators~\cite{Craft:2022kdo,Holguin:2022epo}. Second, it is interesting to extend celestial blocks in conformal theories~\cite{Kologlu:2019mfz,Chang:2020qpj} to non-conformal cases, which will organize the power corrections in a systematic way that respects Lorentz symmetry. Finally, the light-ray OPE provides a new perspective to explore higher twist effects in the perturbative QCD, which is crucial for the precision calculation of QCD contributions in collider physics and for understanding the new factorization and RG features of general gauge theories.


\acknowledgments

We thank David Simmons-Duffin, Sasha Zhiboedov and Hua Xing Zhu for useful discussions and comments on the manuscript. H.C. are supported by the Natural Science Foundation of China under contract No.~11975200 and No.~12147103.

\bibliography{EEC}
\bibliographystyle{JHEP}




\end{document}